\documentclass[%
reprint,
 amsmath,amssymb,
prb,
]{revtex4-1}

\usepackage{physics}
\usepackage{graphicx}
\usepackage{dcolumn}
\usepackage{bm}
\usepackage{hyperref}
\usepackage[utf8]{inputenc}
\usepackage{mathtools}
\usepackage{float}

\usepackage{amsthm}
\usepackage{tikz}
\usepackage{cleveref}
\usetikzlibrary{arrows.meta, calc, decorations.pathreplacing}

\begin{document}


\title{Many-Qudit Representation for the Travelling Salesman Problem Optimisation}
\author{Vladimir Vargas-Calderón}
\email{vvargasc@unal.edu.co}
\author{Nicolas Parra-A.}
\author{Herbert Vinck-Posada}
\affiliation{Grupo de Superconductividad y Nanotecnología, Departamento de Física\\
 Universidad Nacional de Colombia, Bogotá, Colombia}
\author{Fabio A. González}

\affiliation{MindLab Research Group, Departamento de Ingeniería de Sistemas e Industrial\\
 Universidad Nacional de Colombia, Bogotá, Colombia }

\date{\today}

\begin{abstract}
We present a map from the travelling salesman problem (TSP), a prototypical NP-complete combinatorial optimisation task, to the ground state associated with a system of many-qudits. Conventionally, the TSP is cast into a quadratic unconstrained binary optimisation (QUBO) problem that can be solved on an Ising machine. The corresponding physical system's Hilbert space size is $2^{N^2}$, where $N$ is the number of cities considered in the TSP. Our proposal provides a many-qudit system with a Hilbert space of dimension $2^{N\log_2N}$, which is considerably smaller than the dimension of the Hilbert space of the system resulting from the usual QUBO map. This reduction can yield a significant speedup in quantum and classical computers. We simulate and validate our proposal using variational Monte Carlo with a neural quantum state, solving the TSP in a linear layout for up to almost 100 cities.
\end{abstract}

\maketitle

\section{Introduction}

Combinatorial optimisation problems (COPs) aim to find an optimal configuration from an usually finite but intractable set of configurations~\citep{papadimitriou1998combinatorial}. The travelling salesman problem (TSP) is one of the most famous COPs and attracts plenty of interest from the scientific community. It is easy to state, but hard to solve: given a list of cities and the distances between them, what is the shortest route to visit them all and return to the origin city? This problem has a great number of applications, most notably in operational research \cite{Lenstra,Palhares,CHATTERJEE1996490,BLAND1989125,agatz2018optimization}. The TSP is an NP-hard problem \cite{Korte2008}, meaning that no known classical algorithm can solve it in polynomial time as a function of the number of cities. In fact, the brute-force way to solve this problem is to consider $(N-1)!/2$ possible routes and to choose the shortest route \cite{Fogel1988}. Other efficient heuristic solvers have been built taking advantage of particular topological features of the expected solution \cite{Kohonen1982}, such as having no crossing paths in a TSP defined on an Euclidean plane \cite{Bertagnon_Gavanelli_2020}.

Quantum devices are promising platforms to solve COPs~\citep{ushijima2021} due to two main reasons: there could be solutions that are significantly faster than the best classical algorithm for a specific COP, providing a quantum speedup~\citep{albash2018adiabatic}; and there could be solutions that scale in polynomial time with respect to the COP size, possibly providing a strong quantum speedup~\citep{papageorgiou2013}. Notice that the first reason is concerned with finding solutions of the COP in less time, but the time still scales exponentially with respect to the COP size. Such an advantage from quantum devices over the best classical algorithms for NP-hard problems~\citep{karimi2012investigating} has already been demonstrated for Ising spin-glasses~\citep{barahona1982computational}, searching a marked item within an unstructured database~\citep{roland2002quantum}, among others~\citep{albash2018adiabatic}.

A strategy to solve COP such as TSP in a quantum computer is to map the TSP to a Hamiltonian such that the solution tour can be deduced from the ground state of the corresponding Hamiltonian. Usually, the TSP is cast into a quadratic unconstrained binary optimisation (QUBO) problem, which can be easily mapped to an Ising spin-glass model~\citep{smelyanskiy2012near,someya2016novel,minamisawa2019high,hertz1991introduction,kastner2005prospects,warren2013adapting,lucas2014ising}, taking $N^2$ qubits to solve the TSP for $N$ cities. This means that the Hilbert space of the corresponding Ising spin-glass model is of size $2^{N^2}$.

In this paper, we propose a different map from the TSP to a physical system composed of qudits instead of qubits, which has a corresponding Hilbert space size of $2^{N\log_2 N}$ for $N$ cities. This reduction of the Hilbert space size is expected to facilitate the problem of finding the ground state (which corresponds to the TSP solution) both on quantum and classical computers~\citep{lin1993exact,schollwock2011density,kandala2017hardware}. A future experiment of our proposal on a qudit quantum-chip is expected to be superior to the best classical algorithms, just as Ising machines have proven to be superior to general-purpose classical computers for the TSP on complementary metal-oxide-semiconductor field programmable gate arrays~\citep{minamisawa2019high,yamaoka201524}, on quantum processing units such as the ones developed by D-Wave~\citep{warren2020solving,warren2017small,dwave}, and on other devices such as a nuclear-magnetic-resonance quantum simulator~\citep{chen2011experimental}. We argue that an implementation of our proposal should also be superior to these Ising machines, this time not because of a quantum speedup (since both Ising machines and qudit quantum-chips are quantum machines), but because of the considerable reduction of the Hilbert space size.

The paper is organised as follows. \Cref{sec:tspqubo} presents the TSP and reviews the usual map to a QUBO problem, or equivalently, an Ising spin-glass problem. Then, in~\cref{sec:tspqudits} we construct the many-qudit Hamiltonian whose ground state solves the TSP. Then, in~\cref{sec:validation} we show a validation of our proposal using a state-of-the-art classical algorithm for finding the ground state of a many-body problem, namely, the variational Monte Carlo (VMC) method with a neural quantum state (NQS) as a variational ansatz. In \mbox{\cref{sec:discusion}}, we discuss the advantages and the difficulties of our approach. Finally, we conclude in~\cref{sec:conclusions}.

\section{QUBO formulation of the TSP}\label{sec:tspqubo}

Conventionally, the TSP can be mapped to a QUBO problem, which is then straight-forwardly mapped to an Ising Hamiltonian~\citep{smelyanskiy2012near,someya2016novel,minamisawa2019high,hertz1991introduction,kastner2005prospects,warren2013adapting,lucas2014ising}. In particular, following the explanation by~\citet{smelyanskiy2012near}, we define a binary variable $z_{i\alpha}$ that is 1 if the $i$-th city is the $\alpha$-th location visited in a tour, and is 0 otherwise. 

The length of the tour is $\sum_{i,j,\alpha} d_{i,j} z_{i,\alpha} z_{j,\alpha + 1}$, where $d_{i,j}$ is the distance between the $i$-th an the $j$-th city. We must also impose that $\sum_i z_{i,\alpha} = 1$ for any $\alpha$ and $\sum_\alpha z_{i,\alpha}=1$ for any $i$ to ensure that every city is visited exactly once, creating a so-called valid tour. These constraints, however, are only useful conceptually. They can be rewritten as $(\sum_i z_{i,\alpha} - 1)^2 = 0$ for each $\alpha$ and $(\sum_\alpha z_{i,\alpha} - 1)^2$ for each $i$, so that finding the minimum-length tour of the TSP is equivalent to minimising the quantity
\begin{align}
    \begin{aligned}
    &\sum_{i,j,\alpha} d_{i,j} z_{i,\alpha} z_{j,\alpha + 1} \\
    + &\sum_\alpha\left(\sum_i z_{i,\alpha} - 1\right)^2 + \sum_i\left(\sum_\alpha z_{i,\alpha} - 1\right)^2,
    \end{aligned}\label{eq:qubohamiltonianzvar}
\end{align}
which is a QUBO problem. It is customary to use penalty coefficients, which multiply the last two terms of~\mbox{\cref{eq:qubohamiltonianzvar}}, as explained by~\mbox{\citet{lucas2014ising,tanahashi2019application}}. The reasoning behind adding those penalty coefficients is that the minimisation of~\cref{eq:qubohamiltonianzvar} might get stuck on local minima (with classical optimisation algorithms or quantum annealing setups), and modifying the so-called energy landscape can improve convergence towards the global minimum\footnote{In this work we do not take into account the penalty terms because in our numerical experiments we sample valid tours only}.

If we map the binary variable to a spin/qubit $\sigma$ via $z\mapsto \sigma = 2z-1$, we obtain the expression of an Ising spin-glass Hamiltonian. Moreover, the ground state of the Hamiltonian is the solution of the TSP, and the corresponding ground energy matches the length of the solution tour by construction. This approach takes $N^2$ qubits, meaning that the Hilbert space's size is $2^{N^2}$.

\begin{figure}[H]
    \centering
    \includegraphics[width=0.5\textwidth]{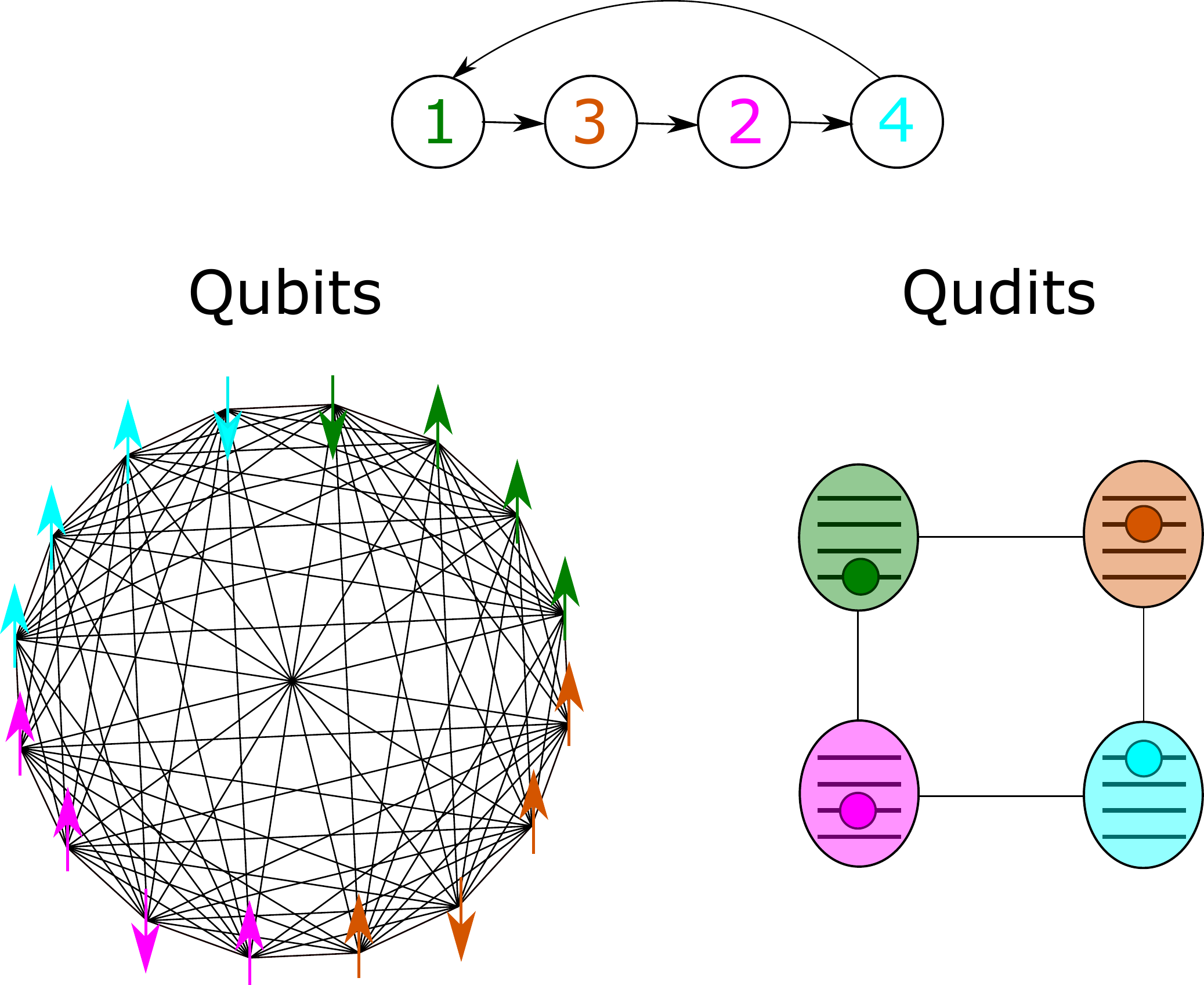}
    \caption{(Color online) Pictorial representation of the TSP map to a system of $4^2$ qubits as a QUBO problem and a system of $4$ $4$-level qudits of a toy-example tour with four cities. The top arrow diagram shows a tour starting at city 1. Colours encode the position of a city in the tour. Lines connecting physical sub-systems represent that they are coupled. The binary variables corresponding to each spin can be read in the order $z_{1,1}, z_{2,1},\ldots, z_{4,1}, z_{1,2},\ldots, z_{4,2},\ldots$, starting from the green spin pointing downwards and in clockwise direction. The last index corresponds to the colour or tour position, whereas the first index corresponds to the city label.}
    \label{fig:mapping}
\end{figure}

Such a representation of the TSP has been experimentally realised in quantum annealing devices~\mbox{\citep{warren2021benchmark,warren2021solving}}, where a set of qubits are sparsely connected between each other, forming a graph where the nodes are qubits, and the edges are couplings between the qubits. However, the Ising Hamiltonian that solves the QUBO problem associated with the TSP requires $2N^2(N-1)$ qubit-qubit couplings (each qubit is connected to $4(N-1)$ other qubits). In ~\cref{fig:mapping} we depict this situation for a four-city tour example, where the QUBO mapping induces qubit-qubit interactions between qubits that represent a single city (connections between qubits of the same colour) and between qubits that represent different cities (connections between qubits of different colours). The interpretation of these interactions becomes cumbersome. Instead, the proposal that we will explain next is more naturally related to representing the cities in a tour and their interactions.

\section{Many-Qudit Formulation of the TSP}\label{sec:tspqudits}

In this work, we propose to use $N$ $N$-level systems or qudits of $N$ dimensions to map the TSP of $N$ cities to the Hamiltonian of a physical system. The corresponding Hilbert space size is $N^N=2^{N\log_2 N}$, which provides an advantage over the qubit proposal. 

We keep the four-city example shown in~\cref{fig:mapping}. We can use four 4-level qudits to encode any tour of 4 cities. Essentially, the tour, which can be described by a string of consecutive cities to be visited $1\to3\to2\to4\to1$, can be encoded in an ordered set of 4 4-level qudits, where the first qudit is in the first-level state, the second qudit is in the third-level state, the third qudit is in the second-level state, and the fourth qudit is in the fourth-level state. This is a more natural representation of the tour than the one-hot encoding into qubits produced by the QUBO representation. In fact, from the string representation of the tour, we can assume a quantum analogue problem where the tour is simply depicted as the pure state $\ket{1}\otimes\ket{3}\otimes\ket{2}\otimes\ket{4}$, which is a tensor product of the 4-level occupation of each of the four qudits.

This is easily generalised to a TSP of $N$ cities. Let $\ket{n_i}$ be the state of the $i$-th qudit. In this setup, the $i$-th qudit occupation refers to the $i$-th visited city. Therefore, any tour can be represented by a vector $(n_1,n_2,\ldots,n_N)$, where $n_i\neq n_j$ for $i\neq j$, which states that the tour begins at city $n_1$, then continues to city $n_2$, and so on, reaching city $n_N$ and then returns to city $n_1$. As discussed, we assume a quantum analogue problem where the tour vector can be represented as a pure state of $N$ qudits, depicted by a tensor product state of the form $\bigotimes_i\ket{n_i}\equiv\ket{n_1,n_2,\ldots, n_N}\equiv\ket{\vb*{n}}$. This allows us to define the Hamiltonian via its element matrices as

\begin{widetext}
\begin{equation}
    \bra{\vb*{n}}\hat{H}\ket{\vb*{n}} = \begin{dcases}
    d_{n_1,n_2} + d_{n_2,n_3} + \ldots + d_{n_N,n_1} & \text{if } (n_1,\ldots,n_N) \text{ is a permutation of } (1,2,\ldots,N), \\
    p & \text{otherwise},
    \end{dcases}
    \label{eq:hamiltonian}
\end{equation}
\end{widetext}
where $p\gg \max\{d_{i,j}\}$ is a term that penalises configurations that do not correspond to valid tours. Such a penalty term can be compared to an effective exclusion principle, where invalid tours cannot exist. Moreover, $\bra{\vb*{n}}\hat{H}\ket{\vb*{n}'} = p$ for $\vb*{n}\neq\vb*{n}'$.

A Hamiltonian of this form may arise from a sum of local Hamiltonians, which are two-qudit operators, whose matrix elements are
\begin{equation}
    \bra{i,j}\hat{D}\ket{\ell, m} = d_{i,j}\delta_{i,\ell}\delta_{j,m} + p'(2-\delta_{i,\ell}-\delta_{j,m}),
    \label{eq:twobody}
\end{equation}
where $\delta_{i,j}$ is the Kronecker delta and $p'\gg \max\{d_{i,j}\}$. Thus, the Hamiltonian of the $N$-qudits system would be
\begin{equation}
    \hat{H} = \hat{D}^{(\mathcal{H}_1\otimes\mathcal{H}_2)} + \hat{D}^{(\mathcal{H}_2\otimes\mathcal{H}_3)} + \ldots + \hat{D}^{(\mathcal{H}_N\otimes\mathcal{H}_1)},\label{eq:twobodyhamiltonian}
\end{equation}
where $\mathcal{H}_i$ is the Hilbert space of the $i$-th qudit and $\hat{D}^{(\mathcal{H}_i\otimes\mathcal{H}_j)}$ is the operator in~\cref{eq:twobody} acting on the space $\mathcal{H}_i\otimes\mathcal{H}_j$. Notice that the Hamiltonian in~\cref{eq:twobodyhamiltonian} is slightly different than the one presented in~\cref{eq:hamiltonian} because the penalty term becomes a collection of penalty terms, depending on how many repeated occupations there are in the state. Again, by construction, any state $\ket{\vb*{n}}$ corresponding to a valid city tour will have an energy equal to the tour distance, which is why minimising the energy yields the ground state of the Hamiltonian in~\cref{eq:twobodyhamiltonian}, which corresponds to the TSP solution.

Not only is the Hilbert space of the qudit system much smaller than the qubit system, but there is also a decrease in the possible physical implementation of the qudit chip, as the graph connecting different qudits becomes highly sparse. In fact, the form of~\mbox{\cref{eq:twobodyhamiltonian}} shows a nearest-neighbour Hamiltonian of a 1D system with periodic boundary conditions, which makes it explicit that only $N$ qudit-qudit couplings are needed (in contrast to $\mathcal{O}(N^3)$ qubit-qubit couplings). Thus, in future qudit-based quantum computers, this problem can be solved using qudits arranged in a ring. Also, each qudit must only be connected to a constant number of 2 other qudits (in contrast to $\mathcal{O}(N)$ for the qubit case).

As a final remark, even though we propose a many-qudit Hamiltonian to solve the TSP, it is possible to map it to a qubit-based computer using binary encoding. Such a map preserves the size of the Hilbert space ($2^{N\log_2N}$), at the cost of not being able to define the TSP as a QUBO problem but as a higher-order binary optimisation (HOBO) problem. Proposals of physical systems that can solve HOBO problems are also starting to flourish, such as the work by~\citet{nikita2021}, where the possibility of controlling $k$-body couplings between the binary nodes of a coherent network is suggested.

\section{Numerical Validation}\label{sec:validation}

In order to validate our proposal, we solve the Hamiltonian in~\cref{eq:twobodyhamiltonian} using a recent and powerful technique for finding the ground state of a many-body physical system. The technique is variational Monte Carlo (VMC) with a variational wavefunction defined by a neural network, also called a neural quantum state (NQS)~\citep{carleo2017solving}. Details of VMC and NQSs are given in~\cref{sec:vmc} and~\cref{sec:nqs}, respectively.

\begin{figure}[b]
    \centering
    \includegraphics[width=0.5\textwidth]{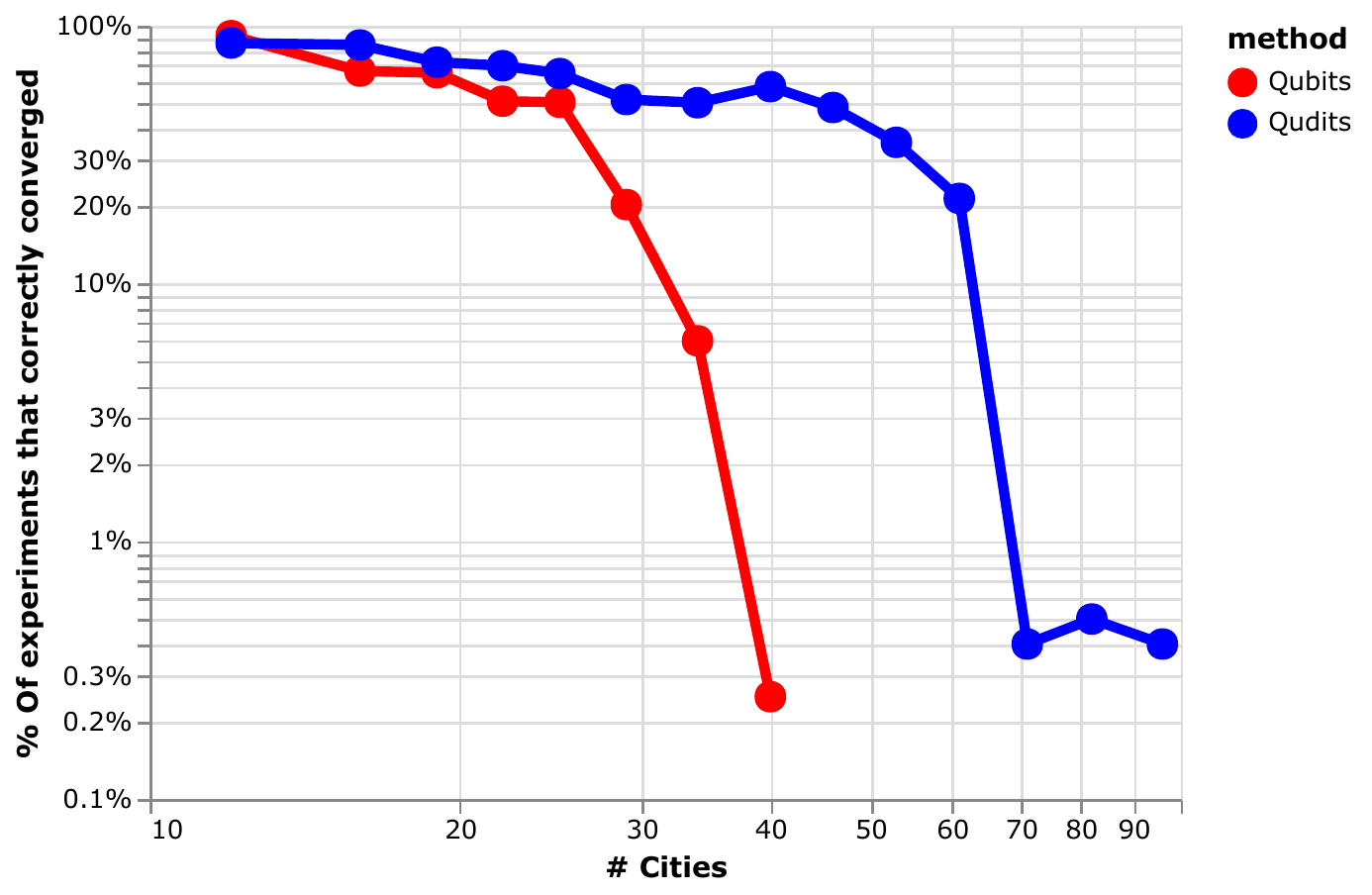}
    \caption{(Color online) Percentage of experiments that converged to the desired solution for the many-qubit (red) versus the many-qudit (blue) representation of the TSP, as a function of the number of cities. The lines are shown to guide the eye only.}
    \label{fig:percentage_convergence}
\end{figure}

\begin{figure*}[t]
    \centering
    \includegraphics[width=0.8 \textwidth]{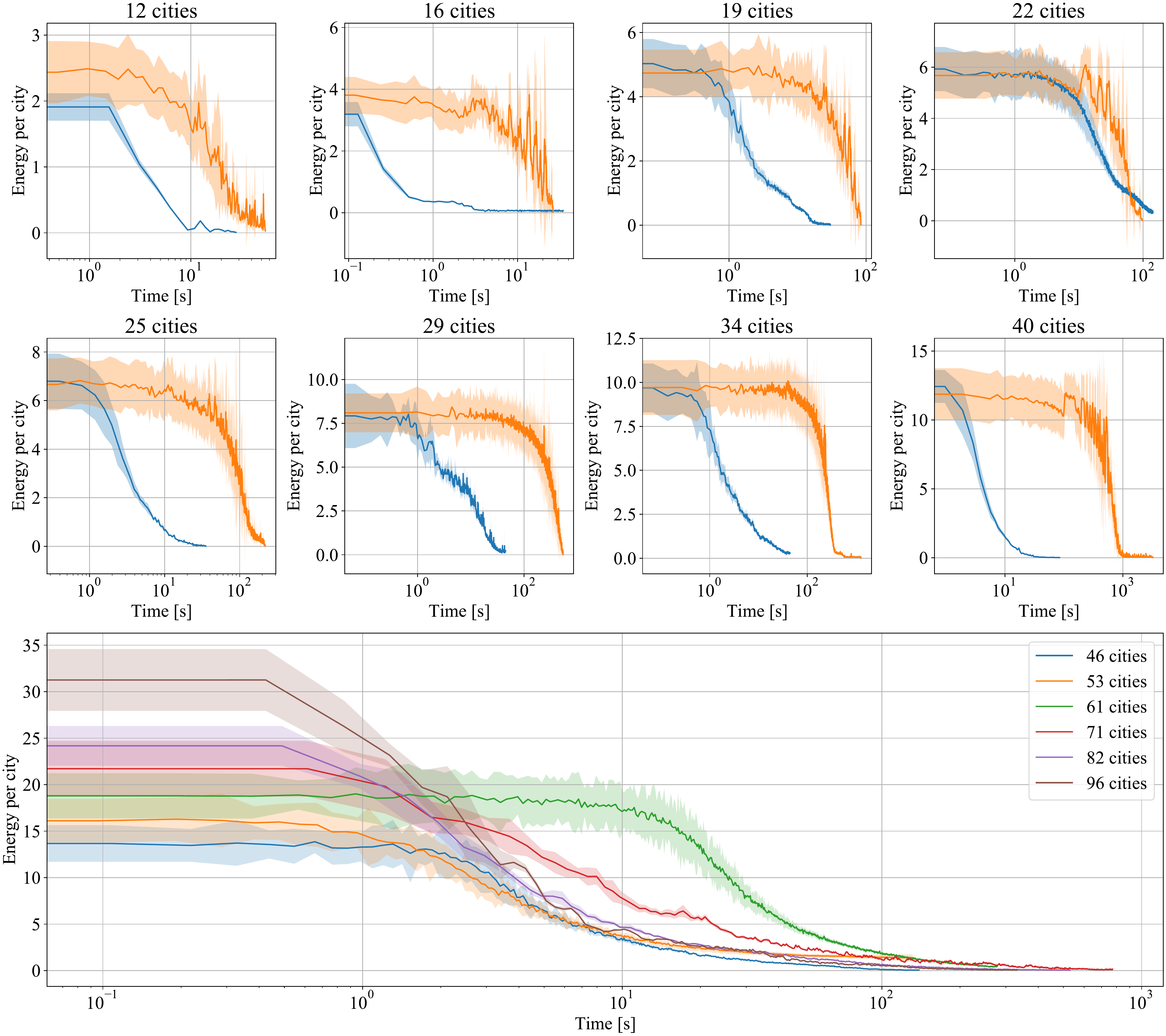}
    \caption{(Color online) Energy convergence as a function of processing time in seconds. The first two rows show the energy convergence of the best experiment for the qubit (orange) and the qudit (blue) representations of the TSP, with respect to the baseline energy, which is $2(N-1)$ for $N$ cities. The bottom panel shows the energy convergence of the best experiments for the qudit representation of the TSP for several other number of cities. In all plots, the shaded areas correspond to 2 standard deviations of the energy of each Metropolis-Hastings sample.}
    \label{fig:convergence}
\end{figure*}

For the sake of illustrating what kind of advantage we can get with our proposal, we perform VMC+NQS experiments on two different setups. The first setup corresponds to the QUBO representation of the TSP, mapped to an Ising Hamiltonian, i.e. a qubit Hamiltonian as shown in~\mbox{\cref{eq:qubohamiltonianzvar}}. For this representation, we use a restricted Boltzmann machine (RBM) as the NQS because it naturally takes as input binary variables. The second setup corresponds to our many-qudit representation of the TSP as shown in~\mbox{\cref{eq:twobodyhamiltonian}}. For this representation, we use a convolutional neural network (CNN) as the NQS because it naturally expresses translational symmetry, which exists in the TSP~\footnote{It does not matter which city is the first one to be visited, the tour length will be the same if we shift the positions of the tour.}. For each setup, we consider a toy-TSP problem, where cities are placed on a line with coordinates $x_n = n$. This class of TSP allows us to quickly benchmark the solutions obtained because the minimum tour length is $2(N-1)$, where $N$ is the total number of cities considered in the city chain. Since the VMC+NQS method has hyper-parameters, we performed 400 experiments with different values of those hyper-parameters for each of the two setups, as it is discussed in detail in~\cref{sec:experimentalsetup}.

\Cref{fig:percentage_convergence} shows the percentage of experiments that correctly converge to the ground energy as a function of the number of cities considered in the linear layout. Interestingly, we see a significant drop in the percentage of experiments that converged to the expected solution around 40 cities for the qubit representation of the TSP (notice that this corresponds to a system with 1600 qubits, which is indeed a very challenging problem). In contrast, the drop is located around 70 cities for the qudit representation. Such a drop indicates how rapidly the TSP solvability decreases with the number of cities, exposing its computational hardness. Moreover, we hypothesise that this drop might be connected to a phase transition of the VMC+NQS algorithm when exposed to the TSP, as this behaviour has previously been seen in other algorithms for the TSP~\citep{gent1996tsp}.

Another essential feature of the experiments carried out is that experiments that converge take less time in the many-qudit representation than in the many-qubit representation. In~\cref{fig:convergence} we exemplify this fact for some number of cities, where the convergence of the best (most accurate and fast) experiments of the qubit (orange) and qudit (blue) representations of the TSP are shown. Not only are the best experiments of the qudit representation better than those of the qubit representation (except in the case of 16 and 22 cities\footnote{Note that the experiments shown in~\cref{fig:convergence} are those that are in a local minimum of the hyper-parameter space.}), but the difference in performance also grows as the number of cities grows.

\section{Discussion}\label{sec:discusion}
The qudit mapping evidences a more straightforward and natural way of solving the TSP than the usual qubit mapping. We have argued in favour of the previous statement from two different perspectives. The first one is that the mathematical structure of the qudit mapping is better suited for the TSP, as there is a natural correspondence between qudit occupation and the city tour position of the travel. The second one corresponds to our classical simulations of both proposals, where we showed that it is indeed simpler to simulate the qudit mapping than the qubit mapping, also yielding better and faster results for the TSP.

The qudit representation of the TSP leads to both a smaller Hilbert space and a sparser graph compared to the qubit mapping. The latter property is relevant in light of quantum chip manufacturing: near-term intermediate-scale quantum (NISQ) devices are being built with qubits connected very sparsely. Thus, implementing dense graph problems (such as the qubit representation of the TSP) requires many more qubits than the physical available in such NISQ chips. We expect that future qudit-based quantum devices will also be sparsely connected. As shown, our qudit mapping proposal matches this sparsity with ease.

Nonetheless, we must highlight that our proposal currently lacks an immediate experimental possible realisation because of several physical difficulties, but there are promising platforms for implementation. First, we must control several levels in a physical system equal to the number of cities. Moreover, occupation-dependent (non-linear) couplings are needed. Qubit-based quantum computers have already been able to reproduce non-linear behaviour, even though this kind of interaction is not native in those computers~\citep{shi2020quantum}. This can be achieved, for instance, with the tuning of boson-mediated interactions~\citep{burd2021quantum}. Also, ultracold atoms in optical lattices such as the one reported by~\citet{meinert2016} show that it is possible to drive the many-body system in such a way that occupation-dependent interactions can be engineered.

Despite the difficulties, the field of qudit-based quantum computers has seen steady progress in recent years towards universal quantum computers~\mbox{\citep{wang2020qudits,wu2020high,imany2019high}}, which are promising and relevant for our proposal. Furthermore, advances in high-dimensional computers have been recently developed, such as the generation of multipartite entangled states in superconducting transmon qutrits \mbox{\citep{cerveralierta2021experimental}} or the generation of 10-level entangled states in a photonic chip \mbox{\citep{Kues2017}}. Taking into account this, our proposal will be feasible once experimental qudit computing matures.

Finally, we remark that the qubit-based representation of the TSP also poses severe problems at present: the number of qubits to solve the TSP grows quadratically with the number of cities, which limits the size of the TSP that can be solved in NISQ devices, primarily due to imperfect control of couplings due to cross-talk or decoherence phenomena. Such issues further aggravate when considering the cubic number of qubit-qubit connections needed to solve the TSP.


\section{Conclusions}\label{sec:conclusions}

We presented a map from the travelling salesman problem (TSP) to the problem of finding the ground state of a many-qudit system\footnote{We provide an open-source library to build a Hamiltonian using both the QUBO and the many-qudit representations of the TSP. The library finds the ground state of the respective Hamiltonian, which coincides with the TSP solution. The library, called Hamiltonian Travelling Salesman Problem (htsp), is available at \url{https://gitlab.com/ml-physics-unal/htsp}. The htsp library is mainly based on the NetKet library~\citep{carleo2019netket}.}. The main feature of this proposal is that the Hilbert space of the system has size $2^{N\log_2N}$, where $N$ is the number of cities in the TSP. This contribution is likely to provide an advantage over the conventional map of the TSP to a QUBO problem, which then can be easily mapped to an Ising spin-glass Hamiltonian. This conventional representation of the TSP has a Hilbert space of size $2^{N^2}$. Therefore, our proposal significantly reduces the size of the Hilbert space.

The main difficulty in building a quantum device able to simulate our many-qudit system is that we require to control occupation-dependent couplings between the qudits, which demands a precise control of non-linear couplings. However, we experimentally validate that our proposal yields correct solutions of the TSP for several configurations of cities on a line on a classical computer with state-of-the-art ground state solvers such as the variational Monte Carlo with neural quantum states as variational wavefunctions.

An interesting perspective is that even though the many-qudit representation is more succinct than the QUBO representation (this is seen from the Hilbert space size, $2^{N\log_2N}=N^N$), it is not the most compact way of encoding all the possible tours. The number of possible tours is of the order of $N!\underset{{N\to\infty}}{\xrightarrow{\hspace*{1cm}}}N^Ne^{-N}\ll N^N$. Thus, a natural question is, which quantum system can support $N!$ states so that the tour configurations can be mapped one-to-one to these states?

Also, although the aim of this paper is not to provide a classical algorithm competitive with the best algorithms for finding TSP solutions, several network architectures can be tested to provide faster solutions. A promising candidate is the class of transformer-like architectures, which have proven to yield interesting results on the TSP~\citep{joshi2020learning} as well as on quantum annealing setups to find the ground state of random Ising spin-glasses~\citep{hibat2021variational,mcnaughton2020boosting}. Furthermore, TSP solvers based on ground state finding can be integrated into meta-heuristic solvers, to solve smaller TSP problems with accuracy.

Finally, we highlight that it remains a challenge to study the induced quantum correlations in the many-qudit system (or the Ising spin-glass corresponding to the QUBO representation of the TSP), as it is not clear how these might affect positively or negatively the solution of the TSP. Furthermore, in realistic quantum devices, the impact of dissipation onto the solution quality of the TSP might be an interesting phenomenon to take into account, especially with dissipative channels such as qubit dephasing or phonon-assisted tunnelling~\citep{berghoff2008resonant}, which are excitation-preserving, thus, maintaining a valid tour configuration.

\bibliography{70060}

\clearpage
\onecolumngrid
\appendix
\renewcommand\thefigure{\thesection.\arabic{figure}}    
\section{Variational Monte Carlo}\label{sec:vmc}

\setcounter{figure}{0}   

In general, the quantum many-body wave function of a physical system can be written as $\ket{\Psi} = \sum_{n_1,n_2,..,n_N} \psi(n_1,n_2,..,n_N)\ket{n_1,n_2,..,n_N} \equiv  \sum_{\vb*{n}}\psi(\vb*{n})\ket{\vb*{n}}$ where $\vb*{n}$ is a set of fermionic or bosonic degrees of freedom. One is usually interested in the ground state, which is a particular $\ket{\Psi}$ (i.e. a particular set of coefficients $\psi(\vb*{n})$) that minimises the expected value of the system's Hamiltonian. Finding the ground state is a QMA problem \cite{kiev_complex} that becomes exponentially hard with the number of degrees of freedom. Variational Monte Carlo (VMC) is a method that tries to solve this problem, by leveraging the well-known variational method~\citep{le2011quantum} in quantum mechanics to quantum mechanical systems with intractable Hilbert spaces~\citep{becca2017quantum}. VMC considers a variational wave function with a set of variational parameters $\vb*{\theta}$, meaning that the coefficients $\psi(\vb*{n})$ are parameterised, i.e. $\psi_{\vb*{\theta}}(\vb*{n})$. Then, as in the variational method, we minimise the expected value of the Hamiltonian $\bra{\Psi_{\vb*{\theta}}}\hat{H}\ket{\Psi_{\vb*{\theta}}}/\bra{\Psi_{\vb*{\theta}}}\ket{\Psi_{\vb*{\theta}}}$ with respect to the variational parameters $\vb*{\theta}$. However, this expectation value is practically impossible to compute, so VMC provides a way to approximate it. By using the completeness relation $\sum_{\vb*{n}}\ket{\vb*{n}}\bra{\vb*{n}} = \hat{1}$,
\begin{equation}
    \expval*{\hat{H}}_{\Psi_{\vb*{\theta}}} = \frac{\sum_{\vb*{n}, \vb*{n}'} \psi^*_{\vb*{\theta}}(\vb*{n}) \bra{\vb*{n}}\hat{H}\ket{\vb*{n}'}\psi_{\vb*{\theta}}(\vb*{n}')}{\sum_{\vb*{n}}\abs{\psi_{\vb*{\theta}}(\vb*{n})}^2}.
\end{equation}

Multiplying the addends in the numerator by $\psi_{\vb*{\theta}}(\vb*{n})/\psi_{\vb*{\theta}}(\vb*{n})$ yields
\begin{equation}
    \expval*{\hat{H}}_{\Psi_{\vb*{\theta}}} = \frac{\sum_{\vb*{n}, \vb*{n}'} \abs{\psi_{\vb*{\theta}}(\vb*{n})}^2 \bra{\vb*{n}}\hat{H}\ket{\vb*{n}'}\frac{\psi_{\vb*{\theta}}(\vb*{n}')}{\psi_{\vb*{\theta}}(\vb*{n})}}{\sum_{\vb*{n}}\abs{\psi_{\vb*{\theta}}(\vb*{n})}^2}.\label{eq:vmcalmost}
\end{equation}
The term $\abs{\psi_{\vb*{\theta}}(\vb*{n})}^2/\sum_{\vb*{n}} \abs{\psi_{\vb*{\theta}}(\vb*{n})}^2$ is the probability $P(\vb*{n})$ of the configuration $\vb*{n}$, which displays the expected value in~\cref{eq:vmcalmost} as an expectation value of a random variable, i.e. it has the form $\expval*{\hat{H}}_{\Psi_{\vb*{\theta}}} = \sum_{\vb*{n}} P_{\vb*{\theta}}(\vb*{n}) f_{\vb*{\theta}}(\hat{H}, \vb*{n})$, which can be approximated by considering only a subset of the configurations $\vb*{n}$ in a sample $\mathcal{M}$. Thus, we have the approximate expectation value of $\hat{H}$ via
\begin{equation}
    \expval*{\hat{H}}  \approx \sum_{\vb*{n} \in\mathcal{M} }\sum_{\vb*{n}'}P_{\vb*{\theta}}(\vb*{n}) \bra{\vb*{n} }\hat{H}\ket{\vb*{n} '} \frac{\psi_{\vb*{\theta}}(\vb*{n'})}{\psi_{\vb*{\theta}}(\vb*{n})},\label{eq:vmcapprox}
\end{equation}
which is a good approximation as long as $\sum_{\vb*{n} \in\mathcal{M} }P_{\vb*{\theta}}(\vb*{n}) \approx 1$. The sum over $\vb*{n}'$ in~\cref{eq:vmcapprox} can be performed exactly because $\hat{H}$ is usually a sparse operator, meaning that for fixed $\vb*{n}$, $\bra{\vb*{n} }\hat{H}\ket{\vb*{n} '} = 0$ for almost every $\ket{\vb*{n} '}$.


Note that VMC is effectively truncating the Hilbert space basis, which is why any method that builds samples $\mathcal{M}$ can be seen as an automatic truncation algorithm. Truncation is necessary most of the times, even with seemingly simple physical systems such as a qubit interacting with a light mode~\citep{tejedor}. In such a system, we can order the states by number of excitations in the system and crop the the states tower at a given number of excitations where the ground-state (or even steady-state for open quantum systems) calculation converges. However, in general, it is not straight-forward to order the basis with a simple criterion.  For this reason, VMC becomes a useful tool, especially for intractable Hilbert spaces, allowing us to discard states that are not relevant for the description of the quantum mechanical system at hand.

In this work we use the Metropolis-Hastings (MH) algorithm \cite{metropolis1953equation} to prepare the sample $\mathcal{M}$ as follows. In the first MH iteration we propose a initial state $\vb*{n_\text{0}}$. In the $j$-th MH iteration, we propose a new state $\vb*{n}_i'$ from $\vb*{n}_i$ using a some update rule. We accept the new state with probability $\abs{\frac{\psi_{\vb*{\theta}}(\vb*{n}_i')}{\psi_{\vb*{\theta}}(\vb*{n}_i)}}^2$. If we accept the new state, then $\vb*{n}_{i+1} \leftarrow \vb*{n}_i'$, else  $\vb*{n}_{i+1} \leftarrow \vb*{n}_i$. We stop iterating after a fixed number of iterations.

\section{Neural Quantum States}\label{sec:nqs}

\begin{figure}[!ht]
    \centering
    \includegraphics[width=0.65 \textwidth]{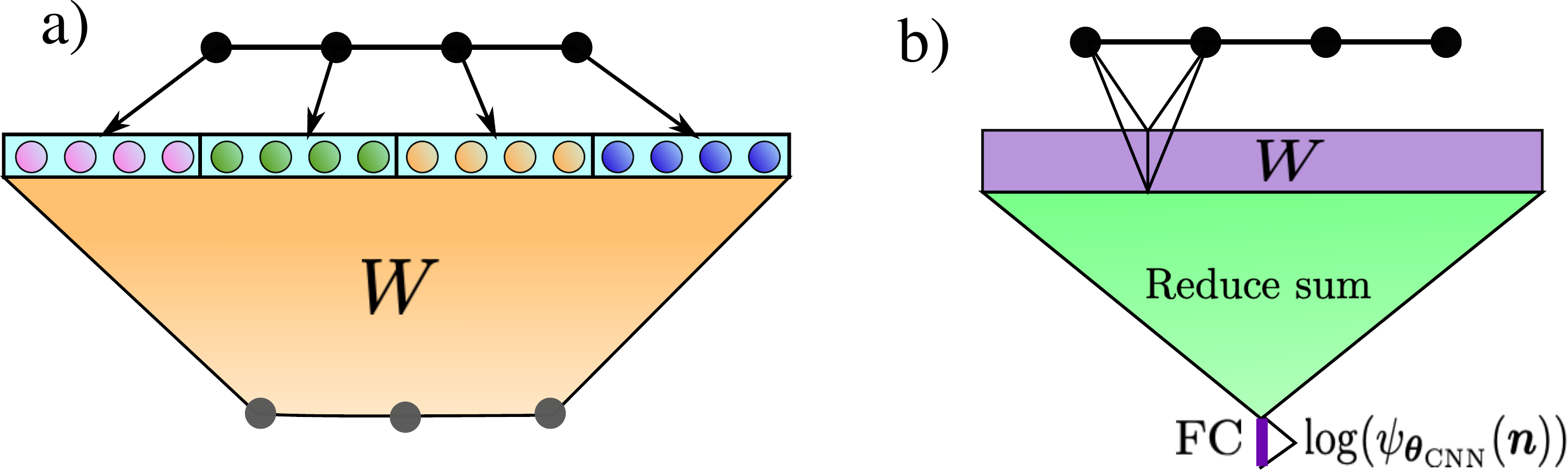}
    \caption{(Color online) Neural quantum states for a) the qubit and b) the qudit representation of the TSP, depicted for four cities. The black line with four nodes represents, in both panels, a tour configuration of 4 cities. a) shows an RBM that uses $4^2$ qubits, which match the visible neurons of the RBM and has three hidden neurons. b) shows a CNN whose output is sum-reduced and fed into a fully connected layer~\citep{yang2020deep}. In both cases, the neural network parameters are complex numbers.}
    \label{fig:waves_functions}
\end{figure}

A neural quantum state $\ket{\Psi_{\vb*{\theta}}}$ defines a wavefunction through the coefficients $\{\psi_{\vb*{\theta}}(\vb*{n})\}$ that result from the evaluation of a neural network with inputs $\vb*{n}$ and parameters $\vb*{\theta}$. In other words, a quantum state is encoded into a neural network, and the wavefunction coefficient corresponding to the city configuration, or city tour $\vb*{n}$ (in the case of the many-qudit system with the Hamiltonian given by~\cref{eq:twobodyhamiltonian}) is obtained by feeding the tour to the neural network. In this work, we consider two different neural networks: one for the qubit representation of the TSP using a spin-glass model, and another one for the many-qudit representation. For the qubit representation, we will use a binary variable neural network that has shown outstanding results in many-body problems, namely the Restricted Boltzmann Machine~\citep{nomura2017,nomura2021helping,vargas2020phase,carleo2017solving} (RBM), and for the qudit representation, we will use a convolutional neural network (CNN).

As in many relevant physical scenarios, tours are unmodified by translational symmetry due to periodic boundary conditions: $\hat{T}\ket{n_1,n_2,\ldots,n_N} = \ket{n_2,\ldots n_N, n_1} = e^{id}\ket{n_1,n_2,\ldots,n_N}$. This also applies to the qubit representation of the TSP~\footnote{There exists a one-to-one correspondence between valid tours in the qubit representation and the qudit representation, namely, a one-hot encoding of the qudit representation yields the qubit representation.}. This is why a natural choice of neural network for the qudit representation of the TSP is a 1-dimensional CNN with periodic padding. As per the qubit representation, translational symmetries can be also imposed to RBMs~\citep{nomura2021helping,nomura2017}.

The specific architectures of both the RBM and the CNN are shown in~\cref{fig:waves_functions}. The coefficients $\psi_{\vb*{\theta}_{\text{RBM}}}(\vb*{\sigma})$ for the qubit representation ($\vb*{\sigma}$ is a vector of $N^2$ qubit values where $\sigma_i\in\{-1, 1\}$) are directly given by the expression~\citep{carleo2017solving}
\begin{align}
      \psi_{\vb*{\theta}_{\text{RBM}}}(\vb*{\sigma})=e^{\sum_{j} a_j\sigma_j}\prod_{\ell=1}^{N_H}2\cosh\left(b_\ell + \sum_j W_{\ell,j}\sigma_j\right),
\end{align}
where $\{\vb*{a}, \vb*{b}, W\}$ are the complex-valued visible bias, hidden bias and connection matrix of the RBM, respectively, and $N_H$ is an hyper-parameter called the number of hidden neurons. On the other hand, the coefficients of the CNN are determined by a 1-dimensional convolutional layer whose output is a matrix with as many rows as cities in the TSP, and as many columns as channels of the convolutional layer (i.e. the number of filters to be applied). More specifically, the output of the convolutional layer is~\citep{yang2020deep}
\begin{equation}
    O_{i, f} = g\left(\sum_{k=1}^K W_{k, f}n_{(i+k)\,\text{mod}\, N} + b_f\right),
\end{equation}
where there are a total of $F$ channels, $g$ is the so-called activation function, which we take to be a rectified-linear unit ($g(x) = \max\{0, x\}$), $K$ is the kernel size of each filter $f$, and $W$ and $\vb*{b}$ are the filter matrix and the bias vector of the convolutional layer. Then, a vector $\vb*{o}$ is obtained through $o_f = \sum_iO_{i,f}$. This vector is an input to a fully-connected layer with one output neuron, which returns $\log(\psi_{\vb*{\theta}_{\text{CNN}}}(\vb*{n}))$.

NQSs tend to induce complicated non-linear dependencies between the parameters $\vb*{\theta}$ and the coefficients $\psi_{\vb*{\theta}}(\vb*{n})$, which is why techniques based on stochastic gradient descent or stochastic reconfiguration are needed to minimise the Hamiltonian expectation value. In particular, we use the Adam optimiser~\citep{kingma2014adam}. This makes VMC an iterative method, where on each Monte Carlo step a sample is built through MH, and parameters $\vb*{\theta}$ are updated. Thus, VMC allows us to navigate the Hilbert space, taking into account only the states that have high probability~\citep{vargas2020phase}. Each MC step will therefore sample a portion of the Hilbert space of the physical system, and will minimise the Hamiltonian. The algorithm converges after repeating MC steps a certain number of times to a local minimum of the energy, which has been empirically shown to coincide with the global minimum of the energy in many VMC+NQS studies~\citep{yang2020deep,nomura2017,nomura2021helping,carleo2017solving,vargas2020phase}.

\section{Experimental Setup}\label{sec:experimentalsetup}

We elaborate a TSP problem that allowed us to ``plant solutions''~\citep{hen2015probing}, which means that there is, by construction, a known ground-state configuration of~\cref{eq:twobodyhamiltonian}. This is useful for benchmarking purposes, as we do not need to use an exact solver to find the correct solution of a TSP problem. We set $N$ cities to be on a straight line. Each city $i$ has a coordinate $x_i=N$. Without loss of generality, we can set the first city in the line at $x=1$ to be the first city in the tour, as this only restricts the salesman to be in a (translational) symmetry sector of the TSP\footnote{The particular layout of our cities-in-a-line problem induces another symmetry resulting in a degeneracy of the ground state, which we do not take into account explicitly by either of the variational wavefunctions. The degeneracy, considering 4 cities, is seen by checking that the tour $1\to2\to3\to4\to1$ is equivalent to $1\to4\to3\to2\to1$ or $1\to2\to4\to3\to1$.}. A solution to the TSP of $N$ cities in this setup is straight-forward to obtain: $(1,2,\ldots,N)$ is a tour that solves the TSP.

The RBM and CNN variational wavefunctions, as well as the VMC, MH and Adam algorithms possess some hyper-parameters which we examine thoroughly. In particular, we have the following hyper-parameters:
\begin{itemize}
    \item RBM. \textit{i)} number of hidden units $N_H$.
    \item CNN. \textit{i)} number of channels $F$; kernel-size of the filter $K$.
    \item MH. \textit{i)} number of Markov chains $N_{\text{MC}}$, which indicates the number of parallel MH processes on a single MC step; \textit{ii)} number of city-swaps $N_S$, which indicates the number of swaps of the MH update rule, meaning that two sites are picked at random from a state $\vb*{n}$ and then are swapped; \textit{iii)} maximum length of swap $\ell_S$, which means that in the update rule, one city is chosen at random, but the other one is also chosen at random, but must be at most $\ell_S$ sites away from the first chosen city.
    \item MC. \textit{i)} sample size $S=|\mathcal{M}|$.
    \item Adam. \textit{i)} learning rate $\alpha$, which controls the amount of change in the neural network's parameters for every MC step.
\end{itemize}

For benchmarking purposes, each Markov chain is initialised in a state built as follows. Using the distance matrix of the cities, take one city at random. Then, pick the city farthest from the first one, and visit it. Repeat until you run out of cities. We use this initialisation method to have a reproducible tour that is certainly not the shortest tour, and allows us to benchmark the VMC+NQS technique.

For a chosen value of number of cities $N$, we comprehensively study the hyper-parameters of our method by performing 400 experiments with different values. The hyper-parameter values for each of the 400 experiments were proposed by the Optuna optimiser~\citep{akiba2019optuna}, which uses sampling and pruning strategies such as the tree-structured Parzen estimator~\citep{bergstra2011algorithms} and the asynchronous successive halving method~\citep{jamieson2016non}, that allow an efficient search of hyper-parameters to optimise an objective function, which in our case is the energy or tour-distance. To avoid experiments that take too much time to complete, or do not show convergence at all, we pruned those runs that did not show improvement in the energy minimisation after 300 MC steps, or that surpassed 5000 seconds for the spin-glass-like model, or 3600 seconds for the many-qudit-like model. A final remark on the MH proposal rule is that we restrict the proposed states to be valid configurations, which is why the penalty terms introduced in~\cref{eq:twobodyhamiltonian} can be ignored.

\end{document}